\documentclass[twocolumn]{article}

\usepackage{threeparttable}
\usepackage{comment}
\usepackage{cite}
\usepackage{times}
\usepackage{amsmath}
\usepackage{amssymb}

\usepackage{graphicx}

\usepackage{color}

\newcommand{\durnum}{m}
\newcommand{\durat}{t}
\newcommand{\durrvar}{T}
\newcommand\T{\rule{0pt}{2.6ex}}       
\newcommand\B{\rule[-1.2ex]{0pt}{0pt}} 

\DeclareMathOperator*{\argmax}{argmax}

\begin{document}

\title{Playtime Measurement with Survival Analysis}

\author{Markus Viljanen,
        Antti Airola,
        Jukka Heikkonen,
        Tapio Pahikkala
        \thanks{M. Viljanen, A. Airola, J. Heikkonen, and T. Pahikkala are with the Department of Information Technology, University of Turku, 20014 Turku, emails: majuvi@utu.fi, ajairo@utu.fi, jukhei@utu.fi, aatapa@utu.fi}
}
\date{}

\maketitle

\begin{abstract}
Maximizing product use is a central goal of many businesses, which makes retention and monetization two central analytics metrics in games. Player retention may refer to various duration variables quantifying product use: total playtime or session playtime are popular research targets, and active playtime is well-suited for subscription games. Such research often has the goal of increasing player retention or conversely decreasing player churn. Survival analysis is a framework of powerful tools well suited
for retention type data. This paper contributes new methods to
game analytics on how playtime can be analyzed using survival
analysis without covariates. Survival and hazard estimates
provide both a visual and an analytic interpretation of the playtime
phenomena as a funnel type nonparametric
estimate. Metrics based on the survival curve can be used to
aggregate this playtime information into a single statistic. Comparison of survival curves
between cohorts provides a scientific AB-test. All these methods work on censored data and enable computation of confidence intervals. This is especially important in time and sample limited data which occurs during game development. Throughout this paper, we illustrate the
application of these methods to real world game development
problems on the Hipster Sheep mobile game.
\end{abstract}

\section{Playtime in games}
\subsection{Why Playtime is Important}

Game analytics is becoming increasingly important in understanding player
behavior \cite{gameanalytics2013}. Widespread adoption of games, internet connectivity and new business models have resulted in data gathering in an unprecedented scale. With increasing availability of data, researches and industry alike are motivated to gain insight into the data through game analytics.

Focal point of analytics is player retention and churn \cite{Seufert2014}. Retention has been used in connection with many related measures and methods aiming to increase the length of product use \cite{Weber2011using,Weber2011modeling,Harrison2012,Harrison2013,Harrison2014,Harrison2015,Debeauvais2014,Debeauvais2015,viljanen2016modeling,viljanen2016useractivity}. Better retention simply means players are engaged with the game for longer. Player churn, meaning players quitting the game either momentarily or definitely, decreases product use and is therefore a counterpart of retention. It has also been extensively researched \cite{Kawale2009,Borbora2011,Borbora2012,Runge2014,Rothenbuehler2015,Hadiji2014,Tamassia2016,sifa2016}. Retention metrics are popular because they are thought to reflect player enjoyment, and increased product use provides increased possibilities for monetization in free-to-play and subscription based games. Game success may be attributed to the process of acquiring new users and retaining these users with effective monetization \cite{Seufert2014}.

Of actual measures that quantify retention in analytics, total playtime is a highly useful overall retention metric \cite{Bauckhage2012,Sifa2014} and session playtime \cite{Henderson2001,Kwok2005,Chambers2005,Feng2007,Tarng2008,Chen2009,Pittman2007,Pittman2010} can be utilized to measure in-game retention. Discrete metrics such as session count \cite{Weber2011using,Weber2011modeling}, progression \cite{Debeauvais2014} and active periods \cite{Kawale2009,Borbora2011,Borbora2012,Runge2014,Rothenbuehler2015,Hadiji2014,Tamassia2016,sifa2016}, have also been used in connection with retention.

In this paper we describe survival analysis methods to
measure retention, with focus on total playtime: this enables
game developers, managers and publishers to better benchmark
the game \cite{gameanalytics2013}. Survival analysis is well-suited for retention
analysis because it is developed specifically for duration data.
Other fields with very long histories where these statistical
methods have become standard include demography \cite{Selvin2008}, reliability engineering and biomedical sciences \cite{Lawless1982}.

The primary reason for its de facto status in many fields is that survival analysis excels with non-normal and censored data, and does not necessitate parametric approaches \cite{Kleinbaum2005}. Playtime exhibits non-normal characteristics: duration is positive, heavily skewed towards zero an often has a long tail. Since in the industry it is often unfeasible to wait until all users have churned to obtain their total playtime, censoring is also present. Furthermore, user retention may not always completely follow popular parametric models \cite{viljanen2016useractivity}, making model-free approaches attractive.
The widespread recognition of survival analysis in fields with
similar data and the demand for scientific analytics suggests that game analytics will benefit from this approach.

\subsection{Related Research}
User behavior in terms of measuring duration data has been researched in game analytics \cite{Bauckhage2012,Sifa2014} and game networking \cite{Henderson2001,Kwok2005,Chambers2005,Feng2007,Tarng2008,Chen2009,Pittman2007,Pittman2010} as a topic of itself, networking often analyzing playtimes along with idle times. Game analytics literature attempts to understand user retention and how the game itself contributes to it. Game networking investigates how network quality and related factors add to user retention and how user activity on the other hand imposes a load on servers which operator might try to mitigate. Total playtime \cite{Bauckhage2012,Sifa2014,Allart2016}, Session playtime \cite{Henderson2001,Kwok2005,Chambers2005,Feng2007,Tarng2008,Chen2009,Pittman2007,Pittman2010} and Session interarrival time \cite{Henderson2001,Kwok2005,Feng2007} or idle time, have been popular measurements. Session is commonly defined as a duration of continuous play \cite{Henderson2001,Kwok2005,Chambers2005,Feng2007,Tarng2008,Chen2009,Pittman2007,Pittman2010} but has also been used to refer to a completed match \cite{Weber2011using,Weber2011modeling,Harrison2012,Harrison2013,Harrison2014,Harrison2015}. With long-term games, popular retention measurements are subscription time \cite{Kawale2009,Borbora2011,Borbora2012} or active periods over calendar time \cite{Runge2014,Rothenbuehler2015,Hadiji2014,Tamassia2016,sifa2016}, possibly combined \cite{Borbora2011}. Session counts \cite{Weber2011using,Weber2011modeling} and progression \cite{Debeauvais2014} are also instances of user retention which are on an abstract level equivalent to a discrete duration variable.

Playtime has been analyzed empirically in many studies with some fitting a parametric playtime distribution. A notable example is the study of total playtimes \cite{Sifa2014} in over 3000 Steam games totaling 6 million players which utilized the Weibull distribution for archetype analysis. Research with parametric models has often investigated exponential \cite{Henderson2001,Kwok2005,Isaksen2015}, Weibull \cite{viljanen2016modeling,viljanen2016useractivity,Chambers2005,Feng2007,Pittman2010,Bauckhage2012,Sifa2014}, Gamma \cite{Bauckhage2012}, log-logistic \cite{viljanen2016useractivity}, log-normal \cite{Pittman2010,Bauckhage2012,Allart2016} and Pareto-type \cite{viljanen2016useractivity,Henderson2001,Kwok2005,Pittman2007} distributions. 

Survival analysis is in the early stages of being applied to game analytics. Some studies have used tools that are central in survival analysis, such as the survival curve of playtimes or the churn rate \cite{Chambers2005,Feng2007,Tarng2008,Chen2009}, where notable is the use of Kaplan-Meier estimate \cite{Tarng2008,Chen2009} to deal with censored session durations due to short collection times. Studies focusing on survival analysis have researched modeling the user process \cite{viljanen2016modeling}, measuring difficulty with automated playtesting \cite{Isaksen2015} and retention regression over time-varying game features \cite{Allart2016}.

\subsection{Survival Analysis of Playtimes}

In this paper we describe how survival analysis can contribute to playtime analysis. We introduce the following fundamental analyses a game analyst can carry out on data consisting of
observed playtimes, using standard survival analysis software such as R \cite{Moore2016}, SAS and Stata \cite{Kleinbaum2005}, among others
\begin{enumerate}
\item Survival and hazard curves: these two foundational
concepts of survival analysis allow studying both
visually and analytically the rates at which players
churn from the game at different time points. They
enable the analyst to better understand the overall
quality, including strong and weak points of a game.
\item Mean and median provide singular metrics for
characterizing the expected and the typical playtime.
They allow the analyst to aggregate the data to a
single informative number together with
confidence intervals.
\item The log-rank test provides a scientific AB-test by
comparing the survival curves of different groups (e.g. players of different versions of a game).
This allows the analyst to deduce whether the groups are different to a given degree of confidence.
\end{enumerate}

We limit our investigation to survival analysis without covariates, which covers the first chapters in textbooks and is often the starting point for any survival analysis investigation \cite{Kleinbaum2005}.
All the presented methods are accompanied with examples
analyzing total playtime in Tribeflame Ltd.’s mobile game
Hipster Sheep. These methods can be used more generally to
analyze any duration data, for retention important targets could
also be session and subscription durations.

\section{Playtime data}
\subsection{Retention}
The target of survival analysis is a positive duration variable, which is often called ‘time to event’. This may be the person’s lifetime in epidemiology, time to machine failure in reliability engineering and time to disease recurrence in medicine \cite{Lawless1982}. Duration variable may also be discrete, such as lifespan in years or repetitions to failure. Formally, a given population has a set of $\durnum$ duration variables $\{\durat_1,\ldots,\durat_\durnum\}$ with $\durat_i>0$.

In gaming, retention refers to player activity until the event of user churn. Based on the literature surveyed there are several candidate duration variables for retention:
\begin{enumerate}
  \item Total playtime
  \item Session playtime
  \item Total progression
  \item Total active or subscription time (MMORPGs etc.)
\end{enumerate}
Total playtime is the total time spent playing the game, in seconds for example. Session playtime corresponds to the duration of continuous play. Total progression relies on a game-developer’s intuition of how game consumption is transformed into a positive non-decreasing value, a natural example would be levels completed. In games with open-ended goals and long-term gameplay, one may analyze total time active as the calendar time player was engaged in the game world, or total subscription periods such as months until cancellation.

\subsection{Churn and Censoring}

Censoring is very common in survival analysis type data. For example, in medical studies patients may drop out of the study before experiencing the event of interest or the study may have a limited follow-up time which terminates the study before every patient has had the event \cite{Kleinbaum2005}. Such subject is called \emph{censored} and the data is in this case is subject to \emph{right censoring}. The study subjects contribute information, since we know time to failure must be greater than the time of censoring. To deal with this possibility, the set of $\durnum$ duration observations $\durat_i$ is extended with corresponding censoring indicators $\delta_i$, to give a data set $\{(\durat_1,\delta_1),\ldots,(\durat_\durnum,\delta_\durnum)\}$. If $\delta_i=1$, we indicate that the time to event is greater or equal to $\durat_i$, and with $\delta_i=0$ that the event occurred at $\durat_i$.

In successful games with very long playtimes, censoring is often unavoidable because game developers wish to perform analytics before waiting every user to churn. Even in mobile games which display short playtimes, scattering of sessions over calendar time implies a very long actual follow-up may be required. There is a second important challenge which is not found in survival analysis. We can always observe if person contracted a disease or a machine failed, but it is not possible in principle know if player has churned definitely. Player may always return to the game; it is only with the passing of time we increase the confidence this will not happen.

The challenge of detecting churn related to total playtime has been dealt with in the literature using various rules to impute churned and non-churned players: assuming players have churned \cite{Bauckhage2012,Sifa2014} or defining a window of inactivity which implies churn \cite{Runge2014,Rothenbuehler2015}. More sophisticated way of churn detection would use one of the churn prediction algorithms in the literature to predict the censoring label \cite{Hadiji2014}, here churned equals 'event observed' and non-churned 'event censored'. Complex user process models are also able to infer churn \cite{viljanen2016modeling}.

This problem manifests with quits without notification because it is sometimes difficult to say if the user comes back. The problem does not exist if the churn event is observable, these include session end, level failure and subscription cancellation. Nevertheless, none of the current solutions seem perfectly satisfactory as they may add bias depending on the method. Player churn is an extended topic and we further assume that a simple method to impute churn is available. This enables us to focus on the standard methods which are universally applicable and have a strong theoretical foundation \cite{Lawless1982}. We aim to discuss extended methods incorporating churn uncertainty in future research.

\begin{figure}
\centering
\includegraphics[width=0.7\linewidth]{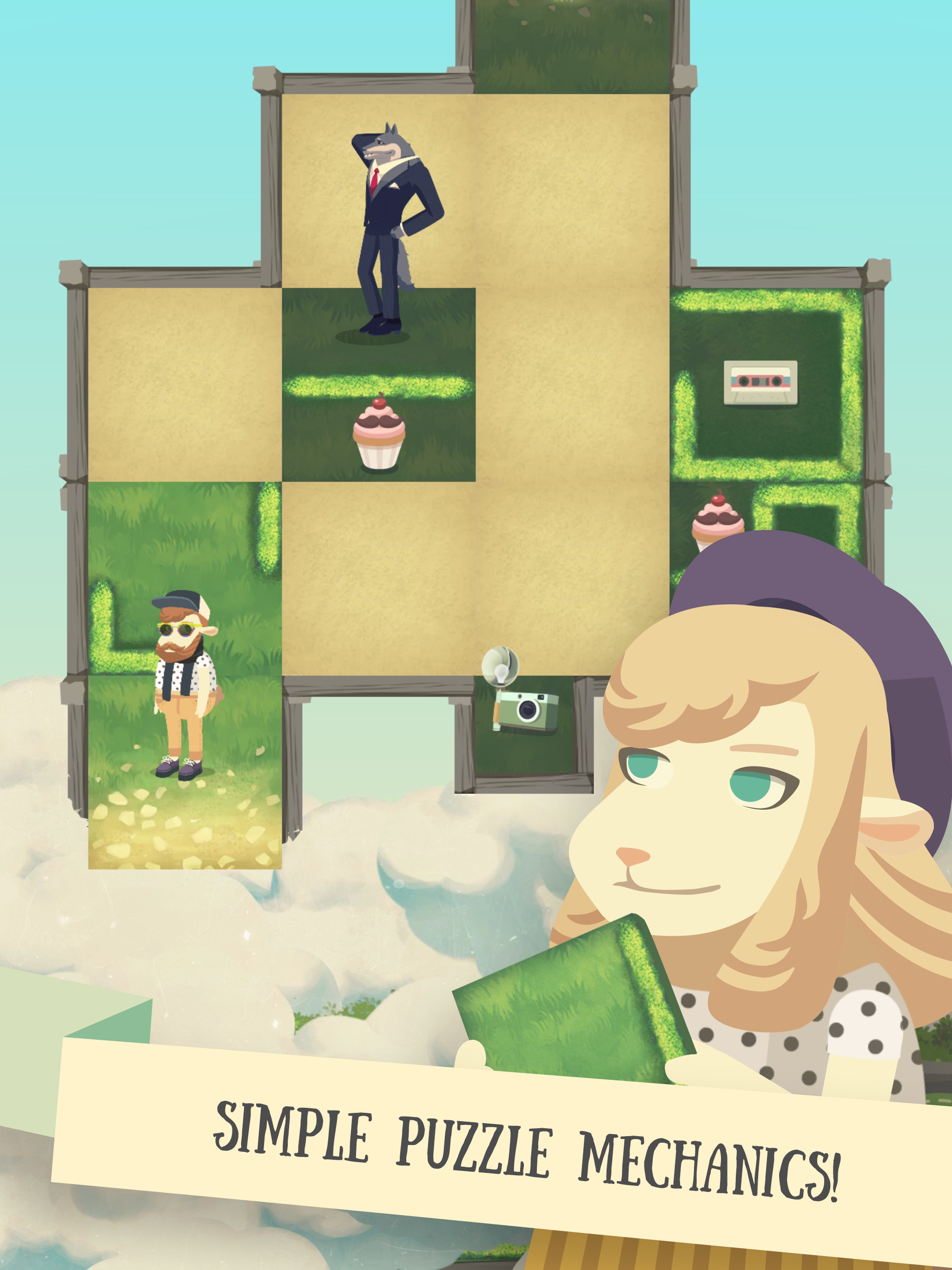}
\caption{Hipster Sheep promotional material also displayed in Google Play, used with permission of Tribeflame Ltd.}
\label{fig:sheepster_screenshot}
\end{figure}

\subsection{Playtime Data Set}

In this study we chose to use total playtimes of an in-development mobile game to illustrate every method with a real world survival analysis application. It is important to keep in mind that methods can be applied equivalently to any game or any duration data motivated by retention measurement.

We use data from Hipster Sheep, in Figure~\ref{fig:sheepster_screenshot}, a commercial grade puzzle game being developed by Tribeflame Ltd. The game is targeted at young adult females and has an artistic theme of making light fun of hipsters through self-irony. The goal is to guide an anthropomorphic sheep through labyrinths on her quest for the next big thing. The game is free-to-play, level-based and uses in-game purchase monetization. Energy mechanics are used to limit the possibility for unlimited free play. Levels combine skill with a hefty dose of luck,
as is common in modern free-to-play games.

\begin{figure}
\centering
\includegraphics[width=0.9\linewidth]{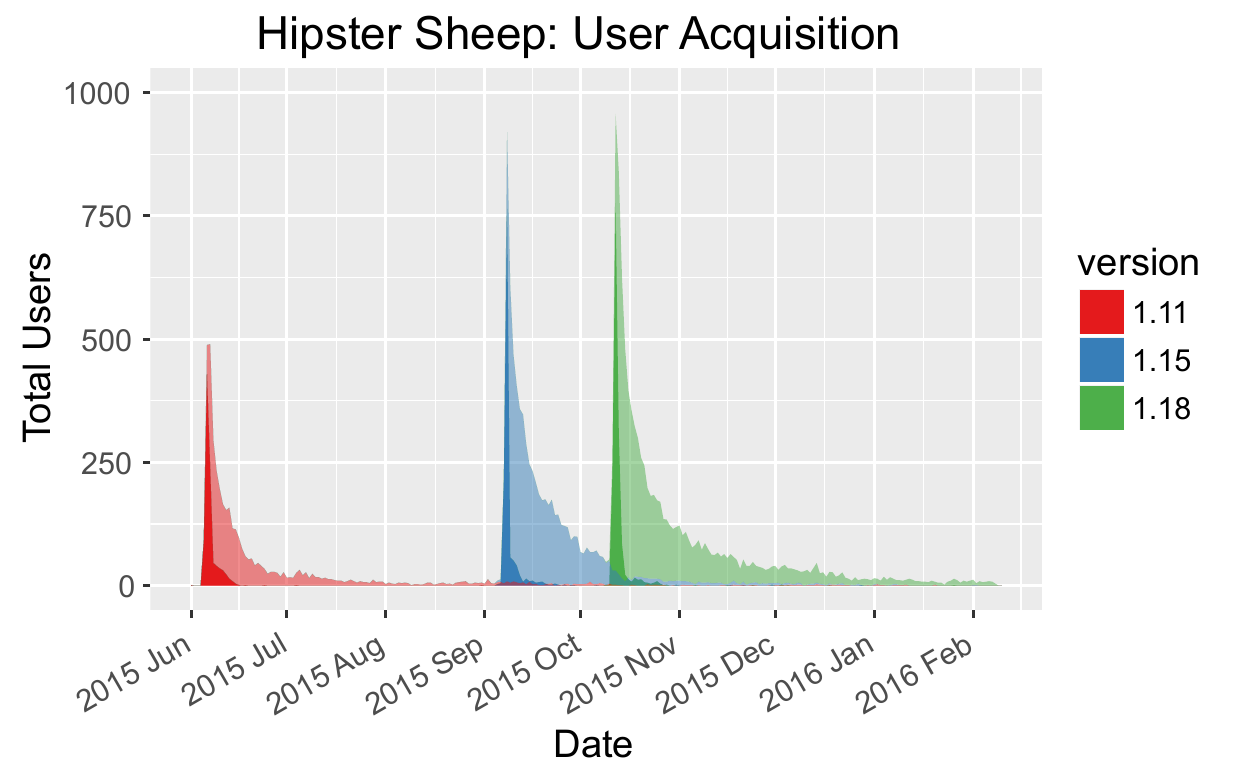}
\caption{User acquisition to test versions 1.11, 1.15 and 1.18. Daily New Users (DNU) are highlighted in dark color and Daily Active Users (DAU) in transparent color. Each acquisition spikes over few days, with resulting user activity diminishing over time.}
\label{fig:sheepster_acquisition}
\end{figure}

During the development, there were three significant user acquisition campaigns for versions 1.11, 1.15 and 1.18, whose purpose was to test the game’s appeal in-between successive development cycles. There are many other version iterations of the game, yet we focus on these versions because they make up the majority of the user base. Users were purchased randomly through advertising in social networks. There was some organic growth where users invited friends or found the game on Google Play, but by and large the data set consists of acquired users.

In Figure~\ref{fig:sheepster_acquisition} we display daily new users (DNU) acquired and resulting daily active users (DAU). In version 1.11 there were a total of 970 users acquired in early June 2015, in version 1.15 a total of 1246 users were acquired early September 2015 and in version 1.18 a total of 1537 new players arrive, mostly mid-October. The three versions hold 3753 players in total. This excludes those players with only one extremely brief session, since this was deduced to be part of 'acquisition phenomena' rather than gameplay; the game takes a dozen seconds to load.

\begin{figure}
\centering
\includegraphics[width=0.9\linewidth]{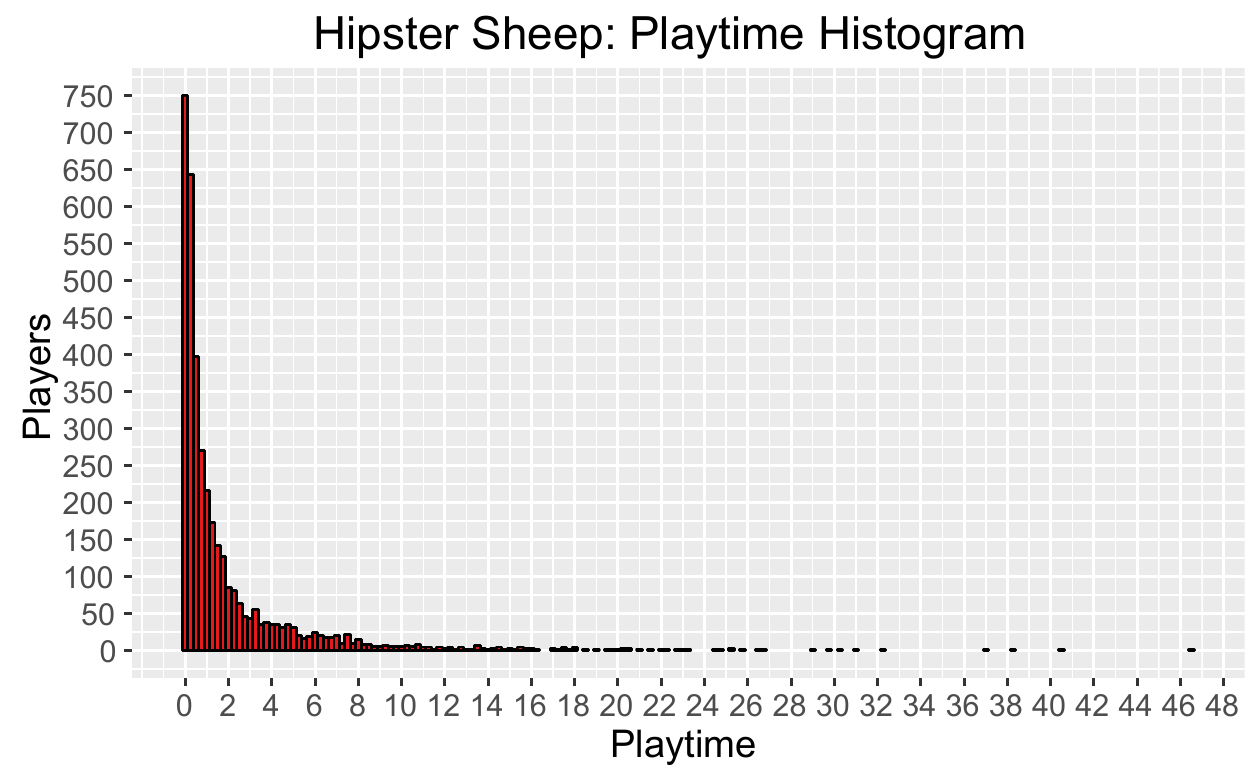}
\caption{A histogram of playtimes for all players in Hipster Sheep, which is one of the most widely utilized tools to measure user behavior.}
\label{fig:hipster_histogram}
\end{figure}

Well-known statistical tools such as histograms, empirical density functions and cumulative density functions are used for duration data across game analytics and networking \cite{Henderson2001,Kwok2005,Chambers2005,Feng2007,Tarng2008,Chen2009,Pittman2007,Pittman2010,Bauckhage2012,Sifa2014,Isaksen2015,Allart2016}. Figure~\ref{fig:hipster_histogram} plots a histogram for comparison to the new methods.

\subsection{Playtime Example Data Set}
The player activity was logged to a database using an in-house logging framework which operates by sending accumulated event packages at brief intervals. The data was processed to compute the total playtime. For almost every player we may confidently say that they churned, but close to the observation limit we decided, following \cite{Runge2014,Rothenbuehler2015}, that players playing 14 days within collection time are not churned. This applied to 1\% of players and led to their playtimes being 'censored'.

To illustrate computations, we randomly sampled a subset of 10 players from version 1.18 for Android. Results are displayed in Table~\ref{table:hipster_10_players} with playtime duration and censoring indicator.

\begin{figure}
\centering
\includegraphics[width=0.9\linewidth]{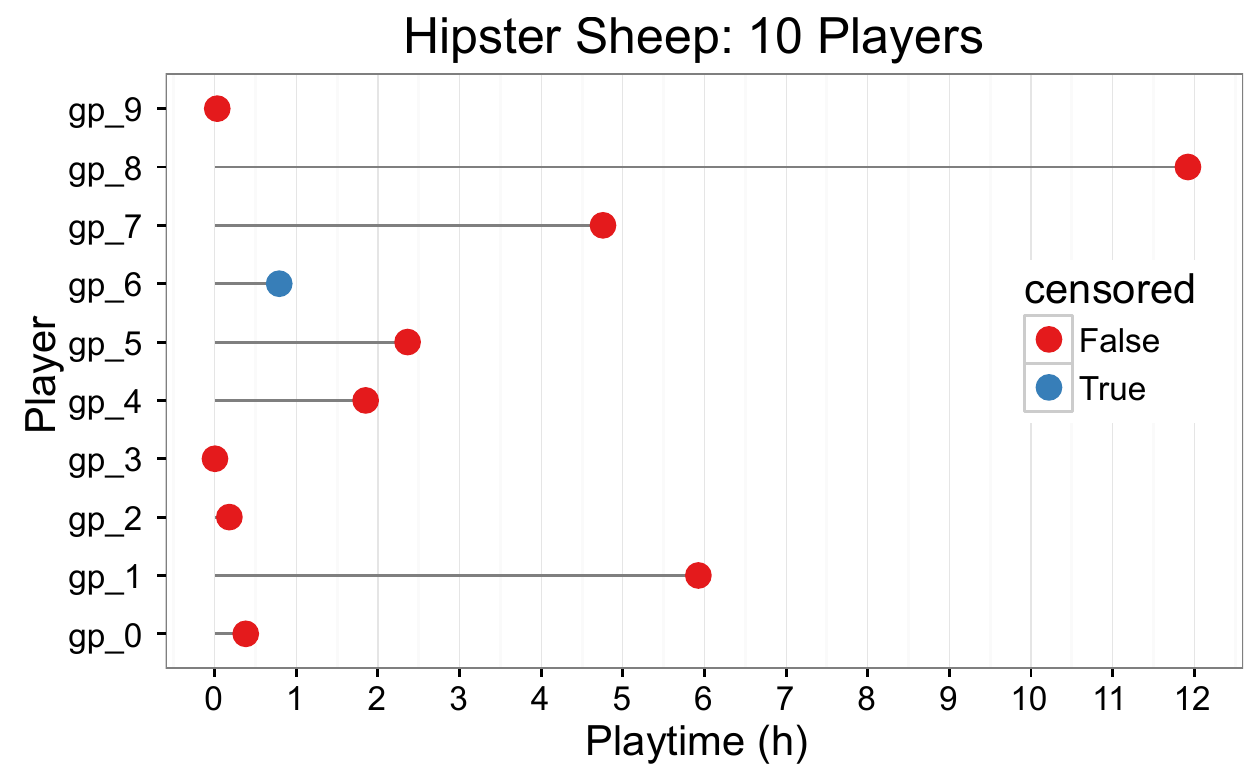}
\caption{A random sample of 10 players from Hipster Sheep version 1.18 for Android. Player with identifier 'gp\_6' had been playing very recently and was determined to be active, whereas others had churned with high likelihood.}
\label{fig:sheepster_sample}
\end{figure}

Figure~\ref{fig:sheepster_sample} visualizes the playtime data of the sample. We see that there is significant early churn, with 40\% churning before 1 hour of gameplay, and a heavy tail with one 12 hour gameplay observation. Other players seem to have more typical 1-6 hour playtimes. One player happened to be censored, but otherwise the sample seems to be quite representative of the population.

\section{Survival Analysis}

\begin{table}
\centering
\caption{Hipster Sheep: 10 Players}
\begin{tabular}{cccc}
\hline
Player & Playtime & Censored & Playtime (hours)\\
\hline
gp0 & 00:22:51 & False & 0.38\\
gp1 & 05:55:32 & False & 5.93\\
gp2 & 00:10:48 & False & 0.18\\
gp3 & 00:00:13 & False & 0.00\\
gp4 & 01:50:59 & False & 1.85\\
gp5 & 02:21:48 & False & 2.36\\
gp6 & 00:47:27 & True & 0.79\\
gp7 & 04:45:25 & False & 4.76\\
gp8 & 11:55:22 & False & 11.92\\
gp9 & 00:01:53 & False & 0.03\\
\hline
\end{tabular}
\label{table:hipster_10_players}
\end{table}

We now begin our investigation on how survival analysis helps to analyze playtime data. In this section, we explain two foundational concepts in survival analysis: the survival curve and the hazard. These concepts enable the analyst to analyze the playtime data both visually and computationally. The survival curve is a natural way to visualize the proportion surviving of a given population, which is why it is typically used in the context of duration data \cite{Kleinbaum2005}. The hazard function is often motivated as the cause for a given survival curve, enabling simpler analysis.

\subsection{The Survival Function}

Suppose $\durnum$ players have playtimes $\{\durat_1,\ldots,\durat_\durnum\}$. Statistically speaking, these playtimes are a sample of a random variable $\durrvar>0$ of the population playtimes and based on the sample we seek to analyze the distribution of $\durrvar$. In the discrete case, where $\durrvar$ is for example playtime rounded to hours, we use a probability mass function (PMF), or the probability of failure at $\durat$: $f(\durat)=\mathbb{P}[\durrvar=\durat]$.
If $\durrvar$ is continuous, we define a probability density function (PDF) instead:
\[
f(\durat) = \lim_{\Delta \durat\rightarrow 0} \frac{\mathbb{P} [\durat\leq\durrvar<\durat+\Delta \durat ]}{\Delta \durat}
\]
Regardless whether we used PMF or PDF, the cumulative density function (CDF) is used to describe the accumulated probability of playtime less or equal to $\durat$. In survival analysis, we often analyze the survival function (SF) \cite{Lawless1982} instead, which gives the probability of having playtime greater than $\durat$:
\begin{align*}
F(\durat)=\int_0^\durat f(u)du=\mathbb{P}[\durrvar\leq t]
\phantom{W}\\
S(t)=1-F(\durat)=\mathbb{P}[\durrvar> t]
\end{align*}
As a complement to the cumulative density function, the survival function is a monotonously decreasing function and has the property that $S(0) = 1$ and $S(\durat) \rightarrow 0$ as $\durat\rightarrow \infty$. Sometimes in practical applications the survival function is not constrained to approach zero, in which case the distribution is improper.

\subsection{The Hazard Function}

\begin{table*}
\centering
\caption{1000 Players Churn Example}
\begin{tabular}{llllll}
Function & Session 1 & Session 2 & Session 3 & Session 4 & \ldots\\
\hline
Retention rate $r(t)$ & 50\% & 80\% & 80\% & 80\% & \ldots\\
Churn rate $h(t)$ & 50\% & 20\% & 20\% & 20\% & \ldots\\
\# Failing $f(t)$ & 500 & 100 & 80 & 64 & \ldots\\
\# Surviving $S(t)$ & 1000 & 500 & 400 & 320 & \ldots\\
\hline
\end{tabular}
\label{table:1000_player_churn}
\end{table*}

Survival analysis distributions are often easiest to understand in terms of geometric decay of players. Since churned players are no longer at risk of churn, it is useful to contemplate constant churn rate acting on the remaining players. We might also have churn rates with high initial churn, and thereon simple constant churn. For example, in Table~\ref{table:1000_player_churn}, 50\% of players play more than one session, but after second session 80\% survive to play session 3, of those 80\% survive to play session 4, etc. Churn in this case refers to the number of players who after session $i$ do not play the next session $i+1$ and survival quantifies the number of players playing more than the $i$th session.

This is formalized in the concept of hazard. For the discrete case, the hazard function \cite{Lawless1982} quantifies the proportion of the remaining players who churn:
\[
h \left (t \right ) =\mathbb{P} \left [\durrvar=t\mid \durrvar\geq t \right ] = \frac{f \left (t \right )}{S \left (t-1 \right )}
\]
For the continuous case, the hazard is the instantaneous failure rate in the remaining player base, motivated as the limit:
\[
h \left (t \right ) = \lim_{\Delta t\rightarrow 0} \frac{\mathbb{P} \left [t\leq \durrvar<t+\Delta t\mid \durrvar\geq t \right ]}{\Delta t} = \frac{f \left (t \right )}{S \left (t \right )}
\]
An important point to note is that continuous hazard is not the probability of failure at time $t$, it can be greater than 1. An approximation in a small interval for the probability of failure is $\mathbb{P} \left [t\leq\durrvar<t+\Delta t\mid\durrvar\geq t \right ] \approx h \left (t \right ) \Delta t$. The relationship of proportional failure probability and rate is analogous to that of PMF and PDF. These two settings can actually be treated together with the Riemann-Stieltjes integral, for more information we refer the reader to \cite{Lawless1982,cook2007}.

\subsection{Connection of Hazard and Survival}

Knowing the hazard or the survival, we can derive the other. In practical applications the hazard is often analyzed for simpler interpretations and the survival curve derived as a function of the hazard. For the discrete case it is easy to see that a product formulation is possible: at point $t$, the survival is the product of the fractions remaining after churn events at $u=1,2\ldots, t$. For example, in Table~\ref{table:1000_player_churn} the survival after session three is: $S \left (3 \right ) = \left (1-50\% \right ) \left (1-20\% \right ) \left (1-20\% \right ) =32\%$, i.e. 320 players.
\[
S \left (t \right ) = \prod_{u=1}^{t} {\left [1- h \left (u \right ) \right ]}
\]
In the continuous case, we take a product integral which works like the Riemann integral in partitioning the domain. Instead of a sum, the result is the limit of a product of terms over the partition, consisting of surviving fractions $1- h \left (t \right ) \Delta t$. In fact, the less well-known product integral  may be written in terms of the Riemann integral by taking the logarithm \cite{cook2007}:
\[
S \left (t \right ) = \prod_{\left(0,t \right]}\left[1- h \left(t \right) dt \right] = \exp{\left[-\int_{0}^{t}{h\left(t \right)dt} \right]}
\]
The integral of the hazard is the cumulative hazard function \cite{Lawless1982}. Its utility is explained in how a proportional change in $S(t)$ corresponds to a linear change in $H(t)$. Taking the logarithm gives the cumulative hazard in terms of the survival function:
\[
H \left (t \right ) = \int_{0}^{t}{h\left(t \right ) dt} =-\log{\left[S \left(t \right ) \right ]}
\]
In the simplest case, the hazard $h \left (t \right )=\lambda$ is homogeneous implying the churn rate is constant over time. This hazard can be used to derive two well-known distributions: geometric distribution for the discrete case and exponential distribution for the continuous case.  Of the common survival distributions listed in Table~\ref{table:survival_distributions}, Weibull is one of the most popular \cite{Harrell2015}, and it also has wide applicability in games \cite{Sifa2014,viljanen2016useractivity,viljanen2016modeling}.

\begin{table*}
\centering
\begin{threeparttable}
\setlength{\tabcolsep}{3pt}
\caption{Simple Survival Type Distributions}
\begin{tabular}{llll}
Function & $S(t)$ & $H(t)$ & $h(t)$\B \\
\hline
Exponential & ${e} ^ {-\lambda t}$ & $\lambda t$ & $\lambda$ \T \\
Weibull & ${e} ^ {- {\left (\lambda t \right )} ^ {\alpha}}$ & ${\left (\lambda t \right )} ^ {\alpha}$ & $\lambda\alpha{\left (\lambda t \right )} ^ {\alpha-1}$ \\
Log-Logistic & $\dfrac{1}{1+(\lambda t)^\alpha}$ & $\log\left[1+(\lambda t)^\alpha\right]$& $\dfrac{\lambda\alpha(\lambda t)^{\alpha-1}}{1+(\lambda t)^\alpha}$ \\
Log-Normal\tnote{a} & $1 - \Phi\left(Z(t)\right)$ & $\log\left[\dfrac{1}{1-\Phi\left(Z(t)\right)}\right]$ & $\dfrac{\phi(Z(t))}{\sigma t \left[1 - \Phi\left(Z(t)\right) \right]}$\B \\
\hline
\end{tabular}
      \begin{tablenotes}
        \item[a] $Z(t)=\frac{ln(t)-\mu}{\sigma}$
      \end{tablenotes}
\label{table:survival_distributions}
\end{threeparttable}
\end{table*}

\section{Playtime Survival}

In this chapter, we utilize the survival and the hazard function
to measure game goodness. We first introduce the theory using
the 10 player sample and then apply the methods to entire game
data. In distribution fitting one has the problem of choosing a
parametric model. However, in survival analysis it is not
actually necessary to guess distributions; the data over the
follow-up time can be used to make model free estimates. These
are called nonparametric methods \cite{cook2007}.

\subsection{Fitting a Survival Model}

Suppose that one has a reason to believe that data follows a parametric model and the hazard or the survival is specified. What next? Fitting a distribution is often done utilizing the maximum likelihood (ML) \cite{Harrell2015}. Specifically, for data set $D$ with durations $\durat_1,\ldots,\durat_\durnum$  and censoring indicators $\delta_1,\ldots,\delta_m$, a PDF/PMF $f(\durat\mid\theta)$ parametrized by $\theta$ is fitted by assuming the observations i.i.d. and finding the parameters $\theta^*$ which maximize the likelihood $L(\mathcal{D},\theta)$ of observed data. 
\[
L \left (\mathcal{D}, \theta \right ) = \prod_{i=1}^{\durnum}{f {\left ({t}_{i} \mid \theta \right )}^{1-\delta_{i}}  S {\left ({t}_{i}\mid \theta \right )}^{{\delta}_{i}}}
\]
\[
\theta^{*} = \argmax_\theta l(\mathcal{D}, \theta)
\]
The logarithm of the likelihood $l(\mathcal{D},\theta)=\log[L(\mathcal{D}, \theta)]$ is taken in practice to avoid numerical errors associated with extremely small quantities. The ML estimate may be found iteratively using optimization algorithms such as Newton-Raphson \cite{cook2007}.

\begin{figure}
\centering
\includegraphics[width=0.99\linewidth]{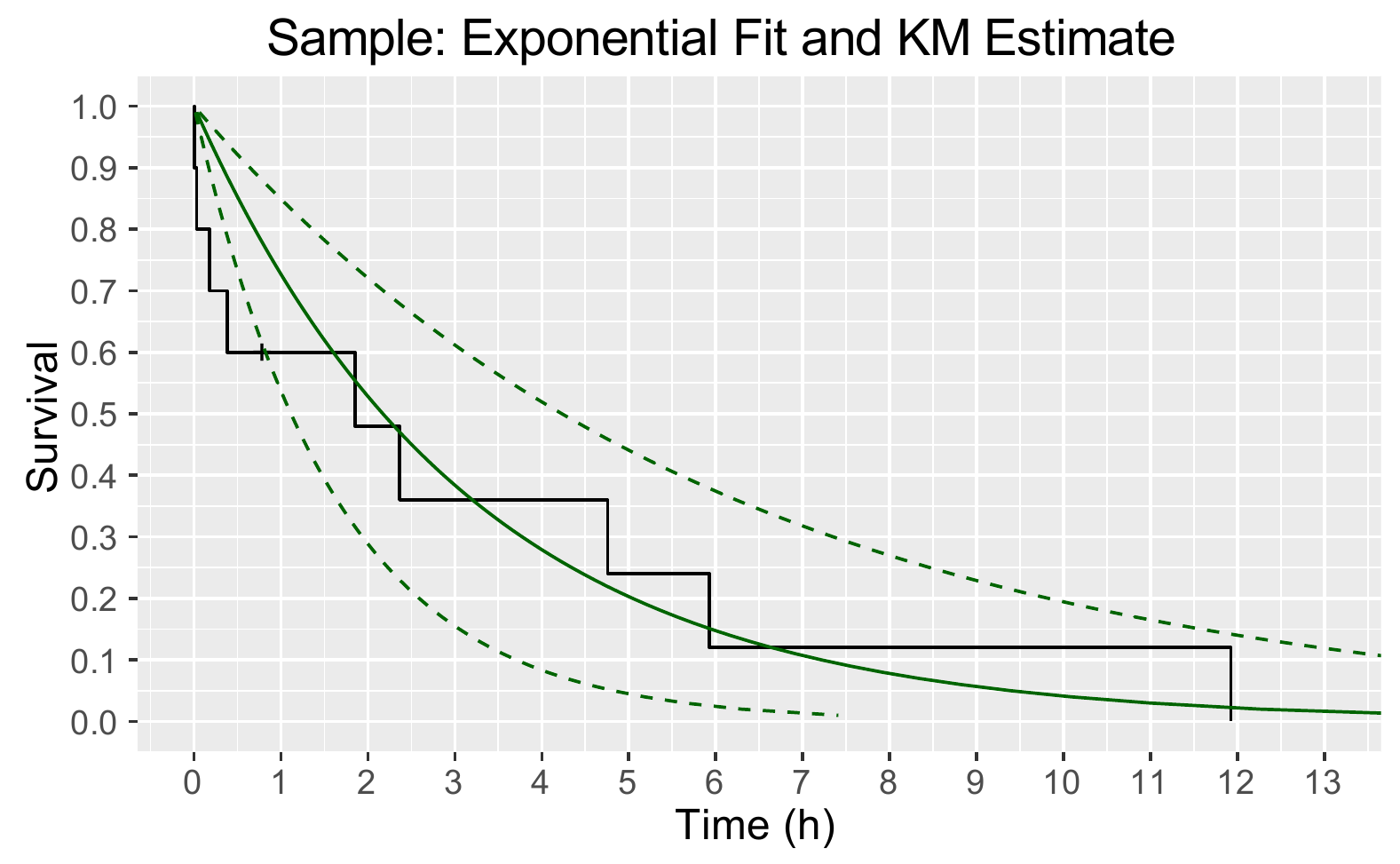}
\caption{Exponential fit (green) $\lambda=0.32$ to the sample of 10 players, with confidence intervals (green, dashed). Contrasted to the KM baseline (black) to be presented later, the early observations may not fit the simple model.}
\label{fig:expofit}
\end{figure}

For example, to fit the exponential distribution in Figure~\ref{fig:expofit}.
one minimizes the likelihood in terms of objective
\[
L(\mathcal{D},\theta)=\prod_{i=1}^\durnum(\lambda e^{-\lambda t_i})^{1-\delta_i}(e^{-\lambda t_i})^{\delta_i} = \lambda^de^{-\lambda R},
\]
where we have defined the number of observed churns $d=\sum_{i=1}^\durnum (1-\delta_i)$ and the total time at risk of churning $R=\sum_{i=1}^\durnum t_i$. The log-likelihood is maximized when the derivative is zero. In this case we can directly find the ML estimate:
\[
l'(D, \lambda^*) =\frac{d}{\lambda^*} - R = 0 \Rightarrow \lambda^*=\frac{d}{R}
\]
Given the survival times in Table~\ref{table:hipster_10_players}, there are 10 players with $d = 9$ churning. The total time at risk is the sum of accumulated playtimes: $R = 0.38 + 5.93 + 0.18 + 0.00 + 1.85 + 2.36 + 0.79_+ + 4.76 + 11.92 + 0.03 = 28.21$ (hours). Therefore, we obtain a failure rate or hazard, of $\lambda =\frac{9 \textnormal{ churns}}{28.21 \textnormal{ h}}= 0.32$ churns/h.

95\% confidence intervals (C.I.) for the parameter $\lambda^*$ may be obtained using the normal approximation $X\pm
1.96\sqrt{\textnormal{Var}[X]}$ where 1.96 is the value of standard normal
distribution such that $\mathbb{P}[-z \leq Z \leq z] = 0.95$. To estimate
confidence intervals for an asymptotically normally distributed
quantity one therefore needs an estimate for its variance.

The variance, or in general the covariance, estimate can be
obtained as the inverse of the observed information, which is
the negative second derivative in this case, or in general is the
negative Hessian \cite{Moore2016}. Since $l''(\mathcal{D},\lambda) = -d/\lambda^2$, the variance estimate can be obtained with a substitution of the maximum likelihood parameter
\[
\textnormal{Var}[\lambda]=\left(-(-d/{\lambda^{*2}})\right)^{-1}=d/R^2.
\]

The failure rate estimate with confidence intervals therefore is $\lambda =
0.32\pm0.21$ churns/h.

\subsection{Estimating Playtime Survival}

Since a chosen parametric model may not always fit the data,
it is often desirable to use an empirical estimate as a benchmark.
If there are no censored observations, it is straightforward to compute the SF empirically as the fraction of playtimes greater than $t$: 
$\widehat{S}(t) = \sum_{i=1}^\durnum \mathbb{I}\left(t_i > t\right)$. However, if there are censored observations, we need to use a Kaplan-Meier estimate.

\begin{figure}
\centering
\includegraphics[width=0.99\linewidth]{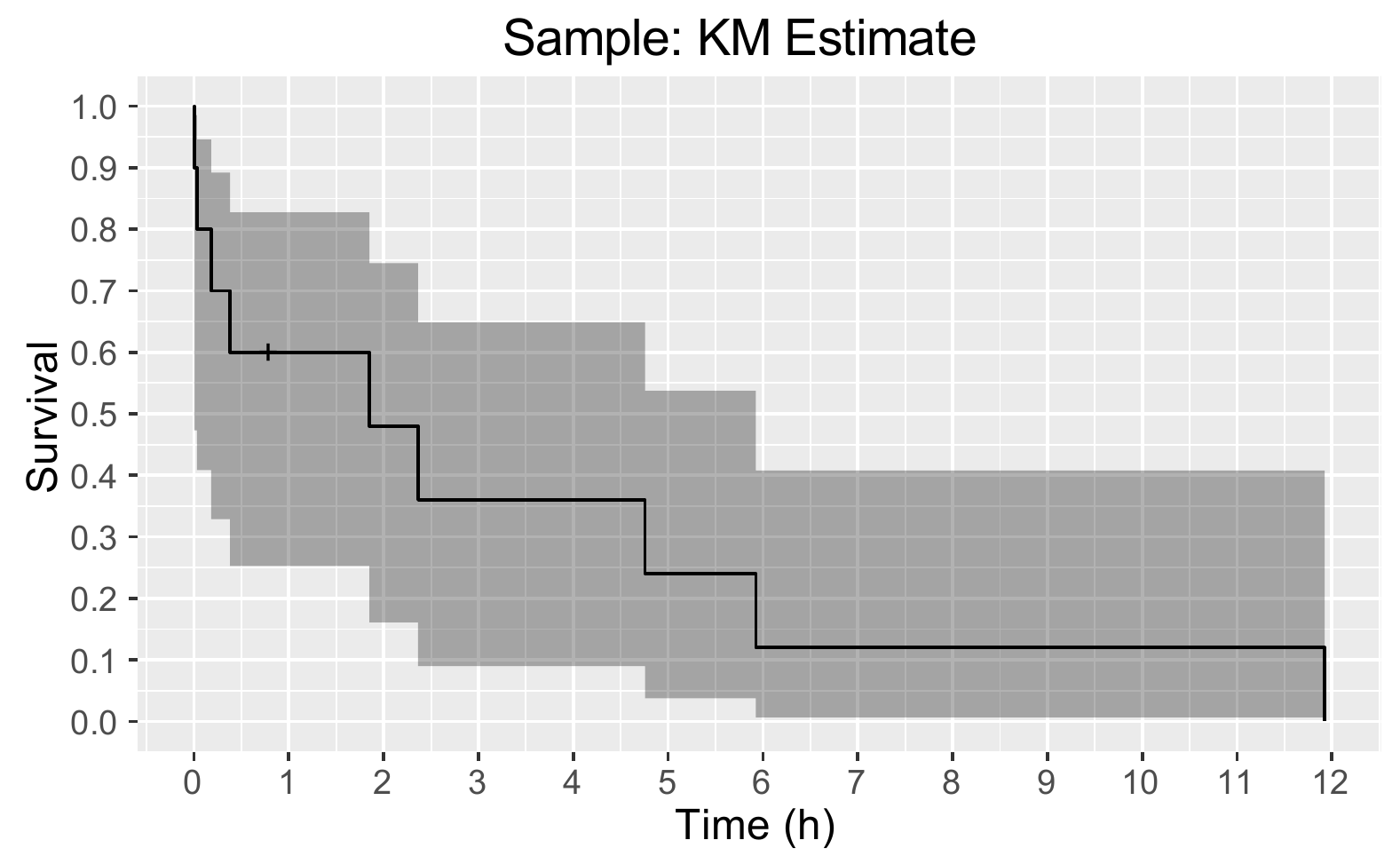}
\caption{Kaplan-Meier estimator (line) with confidence intervals (shaded) for the sample of 10 players. Censored event time at 0.79 is denoted by a small vertical tick. The estimate reaches 0 at the last failure at 11.92 hours.}
\label{fig:sample_KM}
\end{figure}

Given $m$ event times $t_1 , \ldots , t_m$ and at time $t_i$ the number of surviving non-censored, or 'at risk', players $n_i$ and churning players $d_i$, the fraction churning is $q_i = d_i /n_i$ the Kaplan-Meier (KM) product-limit estimator is defined \cite{Kleinbaum2005}:
\[
{\widehat {S}}_{KM} \left (t \right ) = \prod_{{t}_{i} \leq t} {\left (1- \frac{{d}_{i}}{{n}_{i}} \right )}
\]
The estimator is simplest to describe with an example. Table~\ref{table:sample_KM_computation}. and Figure~\ref{fig:sample_KM} show the estimate calcuated for data in Table~\ref{table:hipster_10_players}. At every event time $t_i$, we compute the remaining fraction $1 - d_i/n_i$ and multiply it with the KM-estimate of survival at previous failure time $\widehat{S}_{KM} (t_{i - 1})$ to obtain surviving population $\widehat{S}_{KM}(t_i)$. Note how the one censored event time at $t=0.79$ is not in the table of event times but reduces the risk set at $t=1.85$.

\begin{table}
\centering
\caption{Sample: KM Computation}
\setlength\tabcolsep{3pt}
\begin{tabular}{llllllll}
\hline
Time & At & Churn & Haz. & Cum. &  Surv. & CI.l. & CI.u. \\
(h) & risk      & ${d}_i$ & $\frac{{d}_{i}}{{n}_{i}}$ & Haz. & (KM) & 95\% & 95\% \\
 & ${n}_{i}$ & &  & (NA) & & & \\
\hline
0.00 & 10 & 1 & 0.10 &0.10  & 0.90 & 0.47 & 0.99 \\
0.03 & 9 & 1 & 0.11 &0.21 & 0.80 & 0.41 & 0.95 \\
0.18 & 8 & 1 & 0.13 &0.34 & 0.70 & 0.33 & 0.89 \\
0.38 & 7 & 1 & 0.14 &0.48 & 0.60 & 0.25 & 0.83 \\
1.85 & 5 & 1 & 0.20 &0.68 & 0.48 & 0.16 & 0.75 \\
2.36 & 4 & 1 & 0.25 &0.93 & 0.36 & 0.09 & 0.65 \\
4.76 & 3 & 1 & 0.33 &1.26 & 0.24 & 0.04 & 0.54 \\
5.93 & 2 & 1 & 0.50 &1.76 & 0.12 & 0.01 & 0.41 \\
11.92 & 1 & 1 & 1.00 &2.76 & 0.00 & & \\
\hline
\end{tabular}
\label{table:sample_KM_computation}
\end{table}

\subsection{Estimating Playtime Cumulative Hazard}

An alternative is the Nelson-Aalen (NA) \cite{Kleinbaum2005} estimate of the cumulative hazard given by the sum of fractions churning:
\[
{\widehat {H}}_{NA} \left (t \right ) = \sum_{{t}_{i} \leq t} \frac{d_i}{n_i}
\]

A nonparametric hazard estimate \cite{cook2007} requires smoothing the cumulative hazard step function estimate with kernels, for example. Various kernels exists; popular choices are the uniform kernel, Epanechnikov kernel and Gaussian kernel. Kernel $K\left(t \right )$ is a mass of density concentrated around zero with a total area of one, spread determined by bandwith $b$, which gives a hazard estimate:
\[
\widehat {h} \left (t \right ) = {\frac{1}{b}} \sum_{i=1}^{m} {K \left (\frac{t- t_i}{b} \right ) \frac{d_i}{n_i}}.
\]

\begin{figure}
\centering
\includegraphics[width=0.99\linewidth]{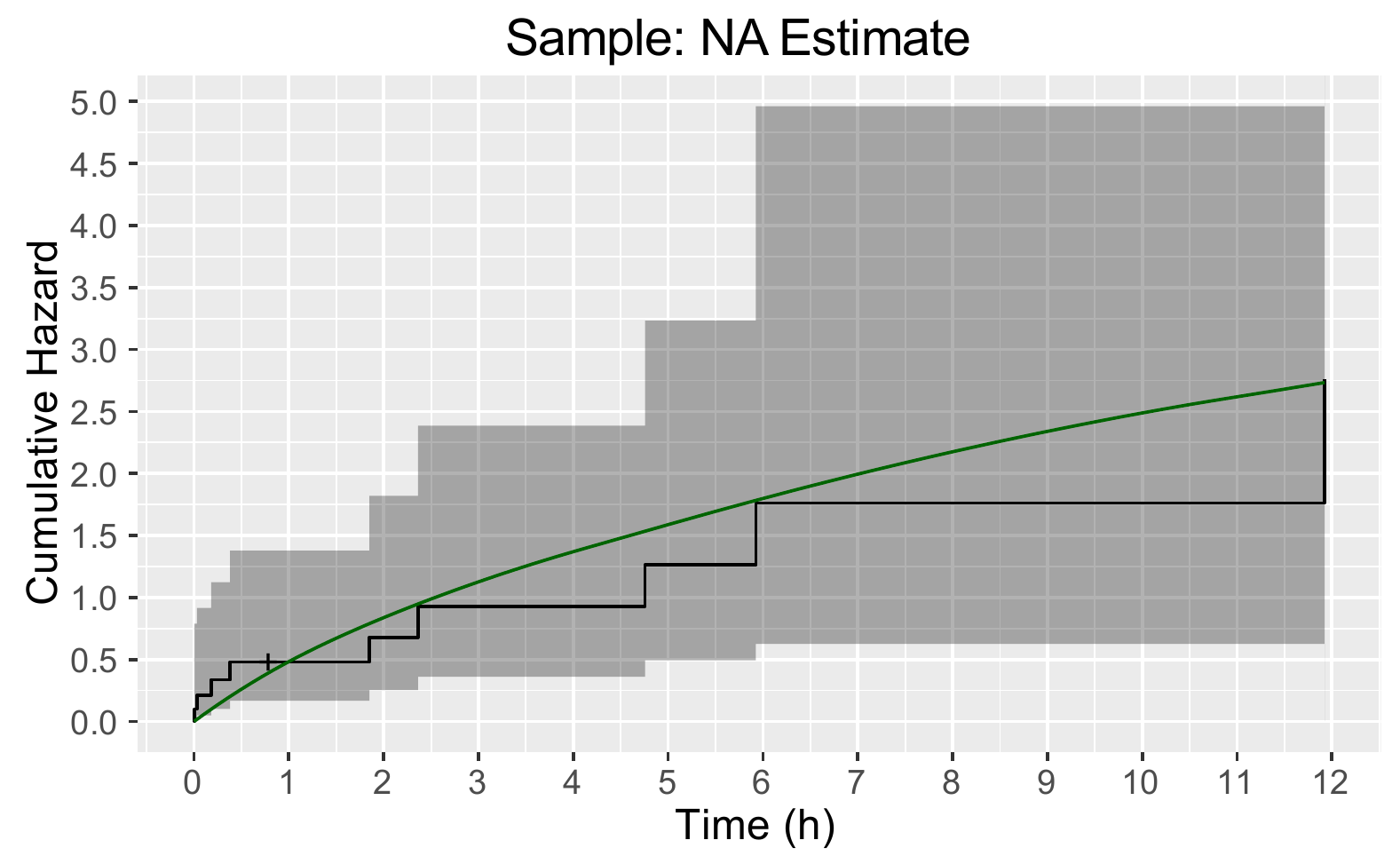}
\caption{Nelson-Aalen estimate (black) with confidence intervals (shaded) and Epanechnikov kernel smoothing (green) for the cumulative hazard in the 10 player sample. The survival corresponding to this estimate never reaches zero.}
\label{fig:sheepster_NA_kernel}
\end{figure}

Of course, using the KM-estimate we can derive a cumulative
hazard extimate by $\widehat{H}_{KM}(t)=-\log[\widehat{S}_{KM}(t)]$. Equivalently for the NA-estimate we have $\widehat{S}_{NA}(t)=\exp{[-\widehat{H}_{NA}(t)]}$. Both estimators are utilized extensively in practice \cite{cook2007}.

It is possible to compute confidence intervals for the KM. The
variance may be approximated with the delta method \cite{cook2007}:
\[
\textnormal{Var}[\widehat{S}_{KM}(t)]=[\widehat{S}_{KM}(t)]^2\prod_{t_i\leq t}\left(\frac{d_i}{n_i(n_i-d_i)}\right)
\]
These variance estimates may extend above and below zero,
which violates survival curve assumptions. A common fix \cite{cook2007}
which provides confidence intervals for NA estimate as well is
to estimate the variance of the log-log transformed estimate:
\begin{align*}
&\textnormal{Var}\left[g[\widehat{S}_{KM}(t)]\right]
\\&\phantom{WW}=\left[\frac{1}{\log[\widehat{S}_{KM}(t)]}\right]^2\prod_{t_i\leq t}\left(\frac{d_i}{n_i(n_i-d_i)}\right),
\end{align*}
where $g(u)=\log[-\log[u]]$. Transforming back with $g^{-1}(u)=\exp[-\exp[u]]$ gives the KM C.I. in Table~\ref{table:sample_KM_computation}.

\subsection{Playtime Survival for Game Data}

The Exponential, Weibull, Log-Normal and Log-logistic
distributions listed in Table~\ref{table:survival_distributions} are common parametric models
for survival data \cite{Lawless1982}. In Figure 8. we have fitted these four models using ML to Hipster Sheep playtimes. We observe three
models with significant model deviations. The exponential
distribution overestimates early survival and underestimates
late survival. The Log-Normal and Log-Logistic distributions
fit short playtimes but have significantly longer tails than
observed in practice. The Weibull distribution appears to have
least model deviation, corroborating the finding that it provides
good approximations to multiple games \cite{Sifa2014}.

\begin{figure}
\centering
\includegraphics[width=0.99\linewidth]{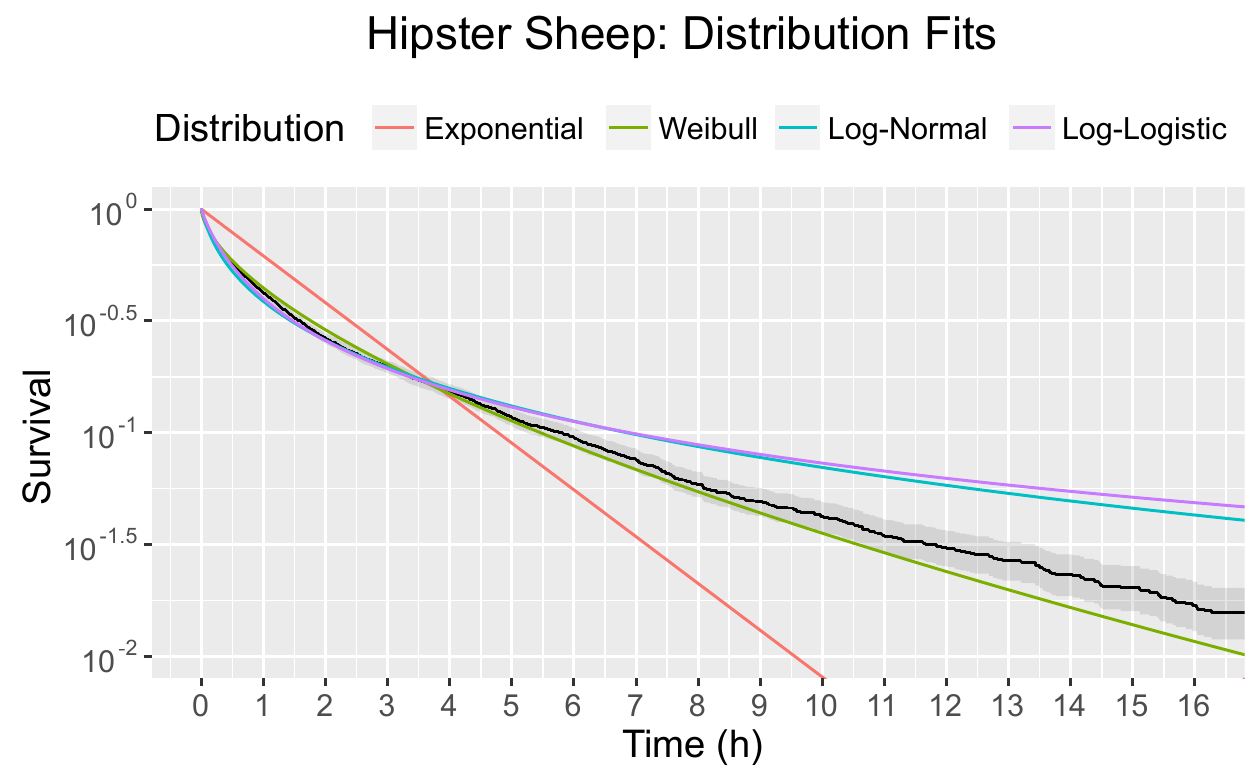}
\caption{Kaplan-Meier estimator and distribution fits for the entire population in
Hipster Sheep. Confidence in the KM estimate is high due to sample size. The
Weibull distribution fits data best with others having significant flaws.}
\label{fig:sheepsters_KM_dist}
\end{figure}

Figure~\ref{fig:sheepsters_KM_dist} demonstrates why the nonparametric Kaplan-Meier
and Nelson-Aalen estimates are popular. Parametric models are
more powerful wherever they describe the data, but when they
do not the results are incorrect. Nonparametric models are
robust to model deviations, in other words they are often the
safe choice, and even in limited data sets they may be sufficient
to describe quantities of interest \cite{Kleinbaum2005}. The provided confidence
intervals are informative in data constrained industry
applications. Since acquiring users costs money \cite{Seufert2014}, a manager
might request a statistically significant user test with fewest
possible users which makes confidence intervals useful.

\section{Playtime Metrics}

Our goal in this section is to explain three important metrics to benchmark the quality of a game. The metrics are motivated by the survival curve: the hazard, the mean playtime and the median playtime. These metrics are simple, easy to measure and it is possible to assess their reliability with confidence intervals.

\subsection{Hazard as a Metric}

In reliability engineering, the failure rate is a key measure of product reliability \cite{Lawless1982}: it provides a profile of how reliability evolves over time. Products may experience early failures due to defective units, the rate may then stabilize to a constant for the period of 'useful life' and go up in the 'worn-out' period. The churn rate provides a similar funnel type visual for games and can be investigated in terms of the early, middle and late-game hazards.
In general, the hazard is an informative
time-dependent metric for the risk of the event in occurring.

In free-to-play games, it is often observed that the failure rate is very high during initial sessions, and stabilizes or steadily continues to decline as most dedicated players remain \cite{viljanen2016useractivity}. In pay-to-pay games with campaigns, one may observe playtimes that are more clustered \cite{Allart2016}. This analysis could prove useful
for game-design as well \cite{Isaksen2015}. In terms of game progression,
good level design should have an approximately uniform churn:
unexpected increases in the churn rate signify flaws and level
specific decreases suggest underutilized improvements.

\begin{figure}
\centering
\includegraphics[width=0.99\linewidth]{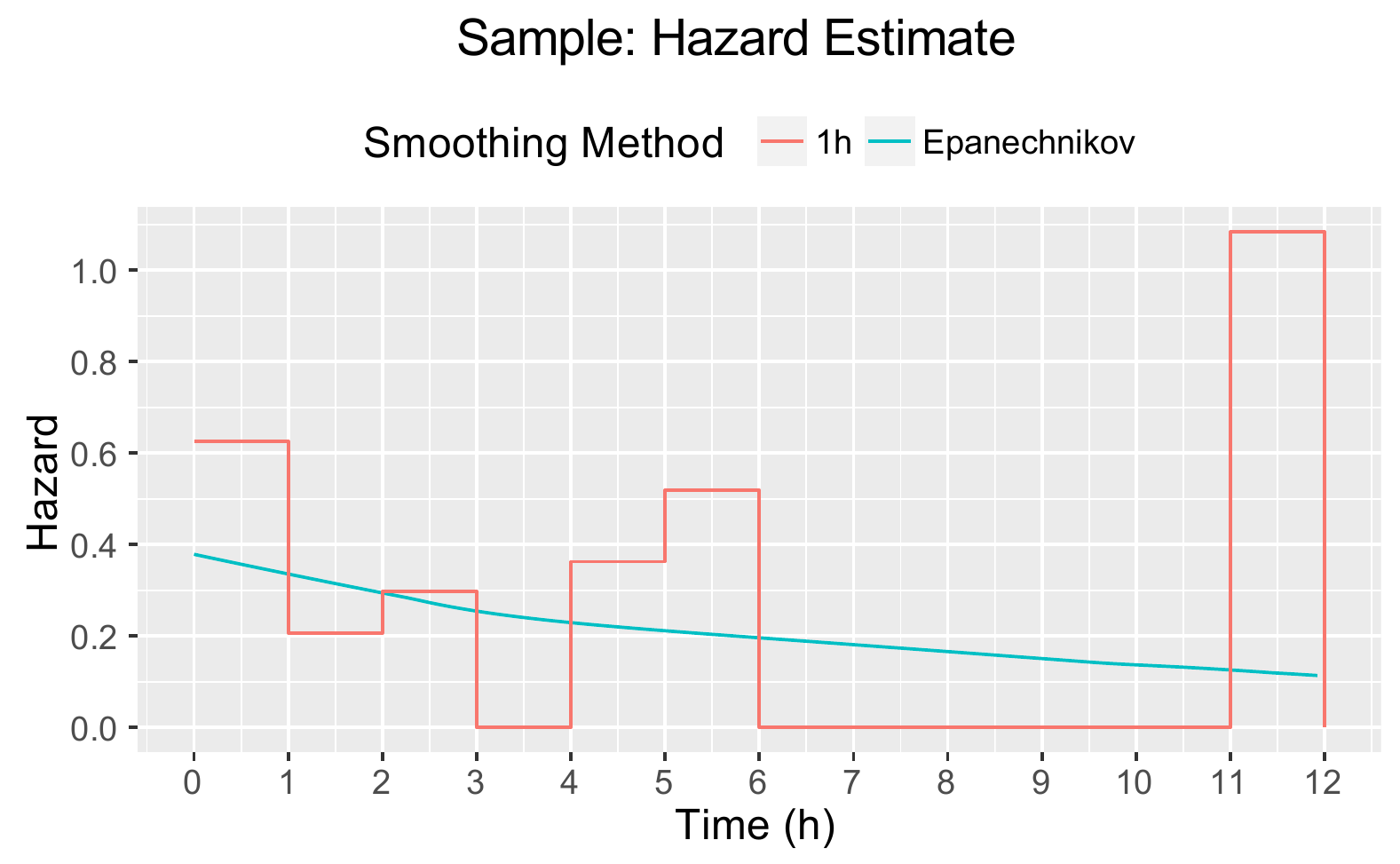}
\caption{Smoothing the 10 player sample with 1 h piecewise exponential rates contrasted to Epanechnikov kernel with $b=9.6$. These bins are too small for this sample, and a higher degree of smoothing appears more informative.}
\label{fig:sheepster_piecewise}
\end{figure}

A major problem with small sample hazard estimation is that
the interpretation could depend on the method chosen. This is
illustrated with the 10 player sample in Figure 9. The piecewise
exponential method has a constant hazard, or exponential
distribution, within given pieces (bins) of the domain. With 1
hour bins, there are $d_{1h} = 4$ churns within the first hour with
total time at risk $T_{1h} = 0.38 + 1_+ + 0.18 + 0.00 + 1_+ + 1_+ +
0.79_+ + 1_+ + 1_+ + 0.03 = 6.38$ h, making the first 1 hour rate
$\lambda_{1h}$ = 4/6.38 = 0.63 churns/h. In bins with no churns the rate
is 0.00, and in the last bin there is 1 churn with a single player
at risk $T_{12h} = 0.92$ h, implying $\lambda_{12}$h = 1.09 churns/h.

\subsection{Mean Playtime as a Metric}

The hazard function is not a single measure, but a function: a set of measures, one for every point in time. Often a single measure is required to benchmark the game. If we assume monetization is proportional to retention it is desirable to use the expected playtime, or the mean, as a singular metric to predict profit which is expected user value minus acquisition cost. For a playtime distribution the mean playtime is defined:
\[
\mathbb{E} \left [T \right ] = \int_{t=0}^{\infty} {tf \left (t \right ) dt}
\]
The mean playtime has a surprising connection to the playtime survival: it is the area under the curve (AUC) \cite{Lawless1982}:
\[
\mathbb{E}\left [T \right ] = \int_{t=0}^{\infty} {S \left (t \right ) dt}
\]
Therefore, to compare two survival curves using a single metric one may compare the area underneath each. This is quite remarkable, since we then have a singular statistic quantifying the goodness of a game. The comparison is well-defined even in cases where the survival curves cross and the ranking is time-dependent. The metric quantifies how much better in additional mean playtime one survival curve is. Visually the shape of the survival curve describes where the additional playtime has been accumulated from: one may have achieved it by decreasing initial churn or increasing long-term retention.

\begin{table}
\centering
\caption{Sample: KM-Based Area Computation}
\setlength\tabcolsep{3pt}
\begin{tabular}{llllll}
\hline
Time & At risk  & Length  & Survival  & Add Area  & Tail Area \\
$t_i$ & $n_i$ & $t_i-t_{i^-}$ & $S_{i-1}$ & $A_i$ & $B_i$ \\
\hline
0 & NA & NA & NA & NA & 3.28 \\
0 & 10 & 0.00 & 1.00 & 0 & 3.27 \\
0.03 & 9 & 0.03 & 0.90 & 0.03 & 3.25 \\
0.18 & 8 & 0.15 & 0.80 & 0.12 & 3.13 \\
0.38 & 7 & 0.20 & 0.70 & 0.14 & 2.99 \\
1.85 & 5 & 1.47 & 0.60 & 0.88 & 2.11 \\
2.36 & 4 & 0.51 & 0.48 & 0.25 & 1.86 \\
4.76 & 3 & 2.39 & 0.36 & 0.86 & 1.00 \\
5.93 & 2 & 1.17 & 0.24 & 0.28 & 0.72 \\
11.92 & 1 & 6.00 & 0.12 & 0.72 & NA \\
\hline
\end{tabular}
\label{table:sample_KM_based_area}
\end{table}

The method of deriving estimate for the mean through area under the survival curve is beneficial because it works with censored observations. Simply ignoring censored observations would lead to downward bias in the estimate. Furthermore, confidence intervals for the mean can be derived utilizing the area. In Table~\ref{table:sample_KM_based_area}. we have computed intervals ${t}_{i} - {t}_{i-1}$ between churn events and how much each event adds to total area ${A}_{i} = {S}_{i-1} \left ({t}_{i} - {t}_{i-1} \right )$. The tail area ${B}_{i} = \sum_{k=i+1}^{m} {{A}_{k}}$, denotes the area remaining in the tail after all areas have been accounted up to $i$. The total area $A= {B}_{0} = \sum_{i=1}^{m} {{A}_{i}}$, which is the expected playtime, can be computed to be 3.28 hours.

95\% confidence intervals using the normal approximation are obtained with $A\pm 1.96 \sqrt {\textnormal{Var} \left [A \right ]}$, in this case $3.28\pm 2.47$ (h). The variance estimate for $A$ can be derived \cite{Selvin2008}:
\[
\textnormal{Var} \left [A \right ] = \sum_{i=1}^{m} {\frac{B_i^2} {{n_i \left ({n} _ {i} -1 \right )}}}
\]

\begin{figure}
\centering
\includegraphics[width=0.99\linewidth]{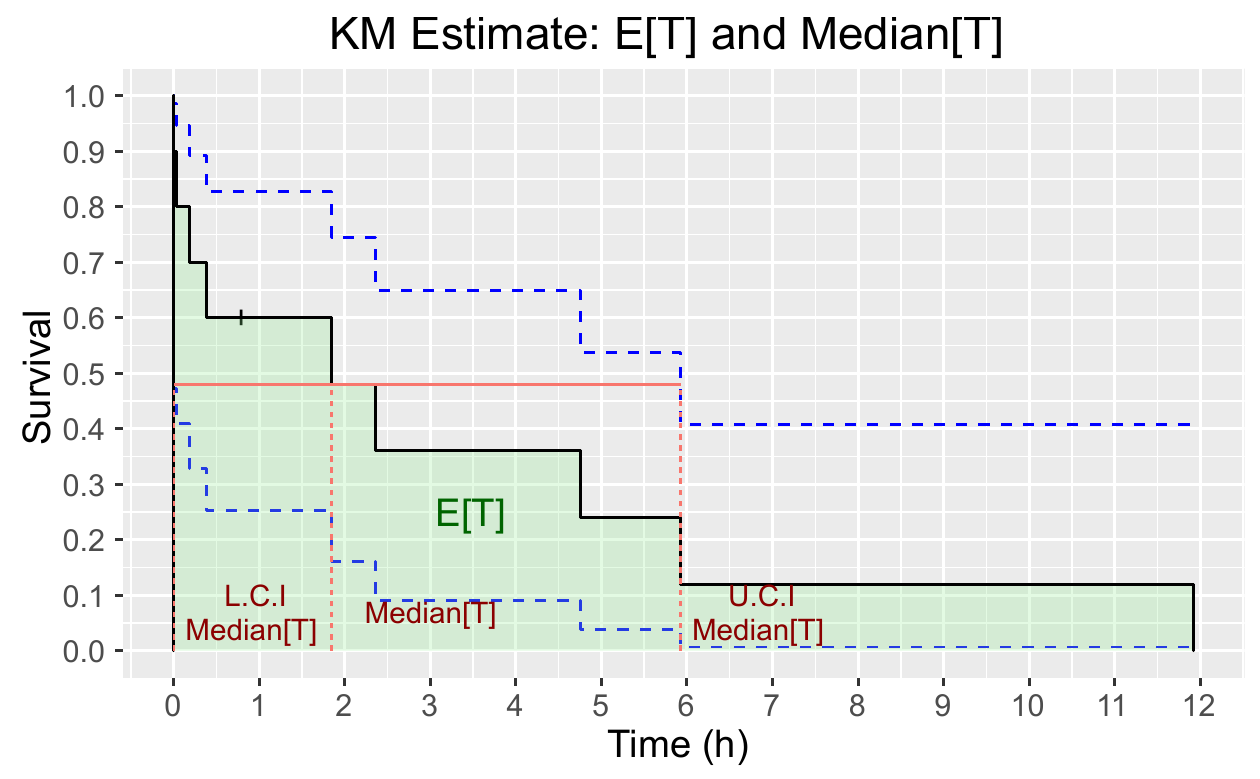}
\caption{Reading E[T] and Median[T] from the KM survival curve. E[T] is the area under the curve (AUC) whereas the curve drops below 0.5 at Median[T] with confidence intervals necessitated by the KM confidence intervals in blue.}
\label{fig:sheepster_metrics}
\end{figure}

\subsection{Median Playtime as a Metric}

The mean playtime is quite informative in many cases, but it may not quantify typical player experience due to many early failures or the presence of a long tail. The median on the other hand, attempts to quantify the 'typical' playtime. It is defined as the point to which half of the players survived:
\[
\textnormal{Median} \left [T \right ] = \min {\left \lbrace t \mid S \left (t \right ) \leq 0.5 \right \rbrace}
\]
In general, one can define arbitrary quantiles for the survival curve. Specifically, the quantiles are defined:
\[
\textnormal{p'th quantile} \left [T \right ] = \min {\left \lbrace t \mid S \left (t \right ) \leq 1-p \right \rbrace}
\]
To compare two survival curves, one may compare the points at which half of players are lost. The game with greater time to lose half of the players is then said to be better, relative to the median. This benchmark may be extended by creating quantile measures, such as a sequence of times $\{ {t}_{10\%} , {t}_{20\%} ,\ldots , {t}_{100\%} \}$ at which $10\%, 20\%, \ldots , 100\%$ of players are lost for each game. These measures give an unequivocal benchmark of short term and long term retention.

The median playtime can be read from the survival curve by finding the earliest t at which the survival drops to equal to or below 0.5. Furthermore, confidence intervals for the median can also be directly read from the pointwise KM-estimate confidence intervals: one draws a vertical line from the median and reads the left (lower) and right (upper) $T$ value at $x$-axis where the line meets the KM C.I. Specifically, we seek lowest and highest value of t such that the following inequality with log-log transform $g \left (u \right ) = \log {\left [- \log {\left [u \right ]} \right ]}$ is satisfied \cite{Moore2016}:
\[
- {z}_{\alpha /2} \leq 
\frac{g \left (\widehat {S} \left (t \right ) \right ) - g \left (0.5 \right )}
{\sqrt {\textnormal{Var} \left [g \left (\widehat {S} \left (t \right ) \right ) \right ]}}\leq {z} _ {\alpha /2}
\]
where $\textnormal{Var} \left [g \left (\widehat {S} \left (t \right ) \right ) \right ]$ was estimated previously to obtain log-log transformed KM confidence intervals. The normal approximation based value is ${z} _ {\alpha/2} =1.96$ for 95\% C.I. The p'th quantile confidence interval may be estimated by substituting $g \left (0.5 \right ) \rightarrow g(p)$ in the inequality. In this case, we obtain a highly uncertain estimate $1.85  \left [0.00\leftrightarrow 5.93 \right ]$  ( h ). 

\subsection{Playtime Metrics for Game Data}

The churn rate provides a very useful time-dependent metric
of game quality. As explained previously, it can be used as a
funnel to quantify strong and weak points of the game. In games
with long-term consumption patterns a simple hazard enables
reliable player lifetime forecasts.

\begin{figure}
\centering
\includegraphics[width=0.99\linewidth]{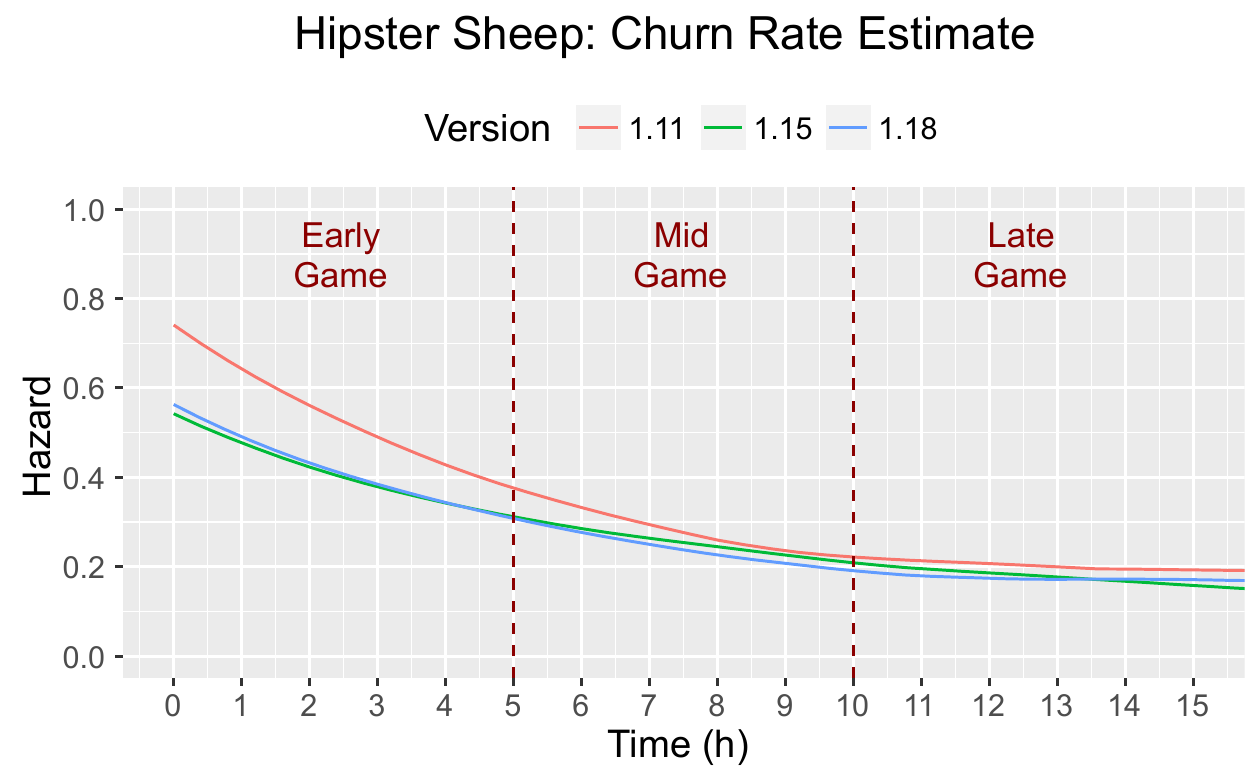}
\caption{The version hazard estimates contrasted with Epanechnikov-kernels using a high degree of smoothing $b=9.6$ and left continuity correction. First version seems uniformly worse, whereas other two appear indistinguishable.}
\label{fig:sheepster_kernel_version}
\end{figure}

In Figure~\ref{fig:sheepster_kernel_version}. we used Epanechnikov-kernels to obtain a
hazard estimate for the three versions in the data set. Since the
playtime is terminated by a churn event, this is a smooth
estimate of the churn rate. We see that version 1.18 churn is quite high
initially at about 0.6 churns/h, and halves to 0.3 churns/h during
first 4 hours, designated as early gameplay. A steady decline
continues as most dedicated players remain in the game and
appears to reach near constant 0.2 churns/h 10 hours onwards.

For the singular metrics that summarize the survival curve,
Table~\ref{table:metrics}. computes the mean and the median with confidence intervals for the three game versions in the data set. One should
note that the mean confidence interval of 1.11 contrasted to
either 1.15 or 1.18 do not overlap, which implies the difference
is significant. However, between 1.15 and 1.18 the confidence
interval is wide enough to contain the other mean estimate. The
quantile metrics, which includes the median, may be read with
confidence intervals from KM estimates in Figure 12.

\begin{table}
\centering
\caption{Hipster Sheep: Playtime Metrics}
\setlength\tabcolsep{3pt}
\begin{tabular}{lllllll}
\hline
Version & Mean & CI.l. & CI.u. & Median &CI.l. & CI.u. \\
& & 95\% & 95\% & & 95\% & 95\% \\
\hline
1.11 & 1.55 & 1.37 & 1.73 & 0.60 & 0.55 & 0.68 \\
1.15 & 2.21 & 2.00 & 2.42 & 0.77 & 0.66 & 0.87 \\
1.18 & 2.41 & 2.18 & 2.64 & 0.77 & 0.66 & 0.86 \\
\hline
\end{tabular}
\label{table:metrics}
\end{table}

The difference between the mean and median as metrics is
clearly visible; whereas random players in 1.18 typically quit
after 0.77 hours, the expected playtime extracted out of random
players is three times larger at 2.41 hours. The presence of both
fickle and dedicated players produces effects which make both
metrics informative from different managerial perspectives.

\section{Playtime Comparison}

\subsection{Comparing Cohort Survival}

Given two survival curves, a test of their difference is often
asked for. Statistical tests can be used for this purpose. These
tests assume that the samples are identically distributed under
null hypothesis, and we obtain evidence which may reject this
conclusion with a given degree of confidence. For example, we
can have two game versions and survival curves produced by
two cohorts consisting of players for each game version. We
then assume the changes had no effect, that the survival is equal,
and the evidence given by players provides a test which may
reject this assumption, leading us to conclude that indeed the
changes affected the game.

While several statistical tests exist, it is useful to be able to compare two survival curves in their entirety. Instead of comparing the means $\textnormal{AUC} \left [{S} _ {1} \left (t \right ) \right ] =\textnormal{AUC} \left [{S} _ {2} \left (t \right ) \right ]$ or the possibility that one is strictly better ${S} _ {1} \left (t \right ) < {S} _ {2} (t)$, we present the log-rank
test \cite{Kleinbaum2005} which tests the assumption that $S_1(t) = S_2(t)$ under censored observations and allows one to use the survival curve
to ascertain the difference. For simplicity we call one of the
cohorts ‘control cohort’ and the other the ‘test cohort’.

\begin{figure}
\centering
\includegraphics[width=0.99\linewidth]{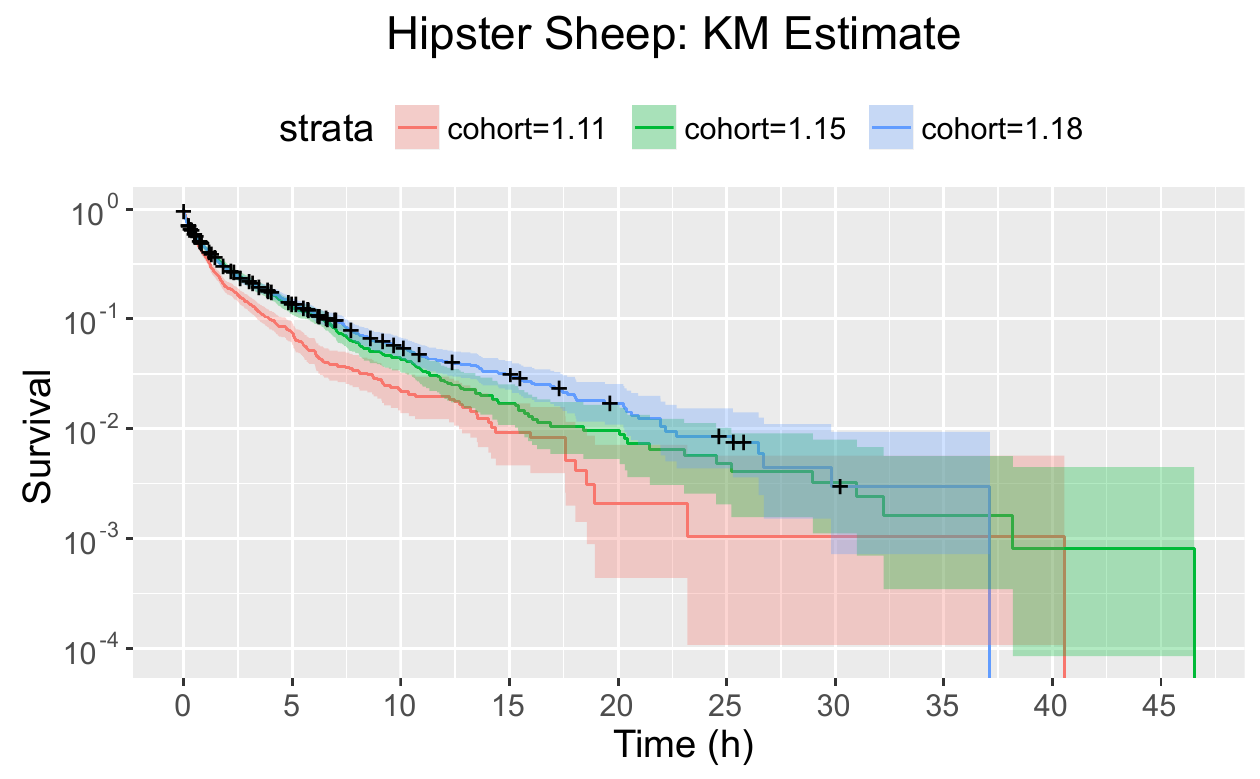}
\caption{Kaplan-Meier estimator for the entire population in Hipster Sheep separated into cohorts by three versions. There is some censoring in the most recent version. Based on the plot, one might think the early version 1.11 is worst, but it is hard to say if the improvement from 1.15 to 1.18 is significant.}
\label{fig:sheepsters_KM_cohorts}
\end{figure}

Suppose that at time $t_i$ there are $n_{0i}$ players with $d_{0i}$ churning in the control cohort and $n_{1i}$ players with $d_{1i}$ churning in the test cohort. Denote $n_i=n_{0i}+n_{1i}$ total players and $d_i=d_{01}+d_{1i}$ total churning. The log-rank test is based on the observation that if the null hypothesis was true, the groups were equal, then given $n_{0i}$, $n_{1i}$, and $d_i$, the number $d_{0i}$ is a sample of a hypergeometric random variable $D_{0i}$:
\[
\mathbb{P} \left (D_{0i}={d} _ {0i} \mid {n} _ {0i} , {n} _ {1i} , {d} _ {i} \right ) = 
\frac{
\binom {{n} _ {0i}} {{d} _ {0i}}  \binom {{n} _ {1i}} {{d} _ {1i}}}
{\binom {{n} _ {i}} {{d} _ {i}} }
\]
The mean and variance of this distribution are \cite{Selvin2008}:
\[
E \left [{D} _ {0i} \right ] = \frac{{n} _ {0i} {d} _ {i}}{ {n} _ {i}}
\phantom{W}
\textnormal{Var} \left [{D} _ {0 i} \right ] = \frac{{n} _ {0 i} {n} _ {1 i} {d} _ {i} \left ({n} _ {i} - {d} _ {i} \right )}{ {n} _ {i} ^ {2} ( {n} _ {i} -1)}
\]
Using these facts, it is possible to construct a linear test statistic based on a score statistic obtained by summing the differences between observed and expected event counts \cite{Selvin2008}:
\[
U= \sum _ {i=1} ^ {m} {\left ({d} _ {0i} -E \left [{D} _ {0i} \right ] \right )}  \phantom{WWW} 
\textnormal{Var} [ U ]= \sum _{i =1} ^ {m} {\textnormal{Var} \left [{D} _ {0 i} \right ]}
\]
A chi-square test statistic allows one to obtain a p-value \cite{Selvin2008}:
\[
\frac{{U} ^ {2}}{ \textnormal{Var} \left [U \right ]} \sim {X} _ {1} ^ {2}
\]

\begin{table*}
\centering
\caption{Statistical Test of Survival Equivalence}
\begin{tabular}{lllll}
Test&${N} _ {\textnormal{control}}$&${N} _ {\textnormal{test}}$&${U} ^ {2} / \textnormal{Var} \left [U \right ]$&$p$-value\\
\hline
${S} _ {1.15} \left (t \right ) = {S} _ {1.11} \left (t \right )$&1246&970&21.4&3.74e-06\\
${S} _ {1.15} \left (t \right ) = {S} _ {1.18} \left (t \right )$&1246&1537&0.4&0.534\\
\hline
\end{tabular}
\label{table:statistical_test}
\end{table*}

To apply this test in a real example, in Table~\ref{table:statistical_test} we have taken the version 1.15 as a control group and compared it to a cohort with version 1.11 and then to 1.18, which are plotted in Figure~\ref{fig:sheepsters_KM_cohorts}. The first test then answers the question whether the version 1.15 was an improvement over 1.11 and the second on whether 1.18 further improved the game. The difference between 1.11/1.15 is highly significant, however unlike what might be visually deduced from Figure~\ref{fig:sheepsters_KM_cohorts}, the 1.15/1.18 difference in the survival tail is completely nonsignificant.

There are modifications to the log-rank test which emphasize different aspects of the survival curve: one may want to weight early or late failures more heavily. For example, the weighted test statistics \cite{Moore2016}
$U= \sum _ {i=1} ^ {m} {{w} _ {i} \left ({d} _ {0i} -E \left [{d} _ {0i} \right ] \right )}$ and
$\textnormal{Var} \left [U \right ] = \sum _ {i=1} ^ {m} {{w} _ {i} \textnormal{Var} \left [{d} _ {0i} \right ]}$
are commonly used with the weight
${w} _ {i} =m \widehat {S} \left ({t} _ {i} \right ) ^ {\rho}$. Setting $\rho=1$ one obtains the Prentice or Peto-Peto modification of the Gehan-Wilcoxon test which places more emphasis on earlier survival differences. In our case, this resulted in $p$-values 0.004 and 0.76 which have the same interpretations.

\subsection{Stratification}

The cohorts may not always be directly comparable. For example, user acquisitions may be conducted with different marketing campaigns or in different countries. Therefore, differences between two versions might really reflect a different underlying composition of players and not changes in behavior due to versions, for example.

To correct for such effects, one needs to adjust for the covariate which is suspected to be an alternate cause for the effects. This is equivalent to testing the null hypothesis ${S} _ {1j} \left (t \right ) = {S} _ {2j} (t)$ across groups $j=1,\ldots,G$. The test is based on computing score statistic ${U}_{g}$ and variance $\textnormal{Var}[U]$ for each group separately and using the test \cite{Moore2016}:
\[
\frac{{\left (\sum _ {g =1} ^ {G} {{U} _ {g}} \right )} ^ {2}}{ \sum _ {g=1} ^ {G} {\textnormal{Var} \left [{U} _ {g} \right ]}} \sim {X} _ {1} ^ {2}\;.
\]
In our case adjusting for country of origin, we obtain p-values 7.88e-06 and 0.608 which does not change the interpretations.

\begin{table*}
\centering
\caption{Applied research cheatsheet}
\begin{tabular}{lll}
\hline
Problem & R function \cite{Moore2016} & R library\\
\hline
Fit nonparametric survival model? & survfit & survival \\
Fit parametric survival model? & survreg & survival \\
Fit nonparametric hazard? & muhaz/pehaz & muhaz\\
Compute mean/median? & print(print.rmean=T/F) & base\\
Compute log-rank test? & survdiff & survival\\
\hline
\end{tabular}
\label{table:cheatsheet}
\end{table*}

\section{Conclusion}

In this study, we demonstrated that survival analysis can be
used to measure retention in games. Positive, skewed and
censored duration data make it a very natural and powerful tool
for this purpose. Duration variables quantifying retention such
as playtime, session time and subscription time, even game
progression, may be analyzed with the methods of survival
analysis. In this study we used a real world game development
example with focus on total playtime.

We presented the basic foundation of survival analysis, which
argued that the phenomena may be analyzed in a simple way
through the churn rate or its complement, the retention rate. The
study focused on three key motivations for survival analysis
based measurement: computing survival curves, deriving
survival metrics and comparing survival data. These methods
contribute towards scientific data analysis by presenting
methods new to game analytics, which are also able to deal with
censoring and utilize statistical significance tests.

For computing survival curves and cumulative hazards, we
presented the Kaplan-Meier and the Nelson-Aalen estimate.
Kernel methods may be used to compute the churn rate and
produce smooth nonparametric survival curves.

For metrics, we discussed how the hazard is an improvement
over using the survival curve as a funnel type estimate. Utilized
widely in reliability engineering, adopting it for game analytics
is especially useful in retention and progression analysis to
detect deviations from the natural pattern of constant rates.
Furthermore, the mean and the median playtime metrics were
derived from the survival curve with confidence intervals.

For survival comparison, we used the log-rank statistical test
to perform a test of the null hypothesis that the survival curves
are equal. The test may be extended to stratify over covariates
and compare multiple cohorts. This method enables scientific
AB-testing of game version quality, for example

The reader may take advantage of Table~\ref{table:cheatsheet} to use the methods for applications. It lists the methods we have presented and the R software functions implementing them.

In summary, survival analysis motivated functions, metrics
and comparisons provide multiple tools to utilize for retention
and progression measurement in game development. We think
that the field has a large potential to contribute to scientific
game analytics and anticipate further research on this topic

\bibliographystyle{ieeetr}
\bibliography{myBibliography}

\end{document}